%% file: main.tex
\documentclass[sigconf]{acmart}

\usepackage{booktabs} 
\usepackage{makecell}
\usepackage{adjustbox}

\begin{document}

\copyrightyear{2018}
\acmYear{2018}
\setcopyright{acmlicensed}
\acmConference[CHASE'18]{CHASE'18:IEEE/ACM 11th International Workshop on Cooperative and Human Aspects of Software}{May 27, 2018}{Gothenburg, Sweden}
\acmBooktitle{CHASE'18: IEEE/ACM 11th International Workshop on Cooperative and Human Aspects of Software, May 27, 2018, Gothenburg, Sweden}
\acmPrice{15.00}
\acmDOI{10.1145/3195836.3195838}
\acmISBN{978-1-4503-5725-8/18/05}

\title{How Do Practitioners Perceive Assurance Cases in Safety-Critical Software Systems?}


\author{Jinghui Cheng}
\affiliation{%
  \institution{Polytechnique Montr\'eal, Montr\'eal, Canada}
}
\email{jinghui.cheng@polymtl.ca}

\author{Micayla Goodrum}
\affiliation{%
  \institution{University of Notre Dame, Notre Dame, USA}
}
\email{Micayla.J.Goodrum.1@nd.edu}

\author{Ronald Metoyer}
\affiliation{%
  \institution{University of Notre Dame, Notre Dame, USA}
}
\email{rmetoyer@nd.edu}

\author{Jane Cleland-Huang}
\affiliation{%
  \institution{University of Notre Dame, Notre Dame, USA}
}
\email{JaneClelandHuang@nd.edu}

\begin{abstract}
Safety-critical software systems are those whose failure or malfunction could result in casualty and/or serious financial loss. In such systems, safety assurance cases (SACs) are an emerging approach that adopts a proactive strategy to produce structuralized safety justifications and arguments. While SACs are recommended in many software-intensive safety-critical domains, the lack of knowledge regarding the practitioners' perspectives on using SACs hinders effective adoption of this approach. To gain such knowledge, we interviewed nine practitioners and safety experts who focused on safety-critical software systems. In general, our participants found the SAC approach beneficial for communication of safety arguments and management of safety issues in a multidisciplinary setting. The challenges they faced when using SACs were primarily associated with (1) a lack of tool support, (2) insufficient process integration, and (3) scarcity of experienced personnel. To overcome those challenges, our participants suggested tactics that focused on creating direct safety arguments. Process and organizational adjustments are also needed to streamline SAC analysis and creation. Finally, our participants emphasized the importance of knowledge sharing about SACs across software-intensive safety-critical domains.
\end{abstract}

%
%



\maketitle

\renewcommand{\shortauthors}{Cheng et al.}

\vspace{-8px}
\input{introduction}
\input{methods}
\input{results}
\input{conclusion}

\section{Acknowledgments}
The work in this paper was partially funded by the US National Science Foundation Grant CCF-1647342.

\bibliographystyle{ACM-Reference-Format}

\end{document}

%% file: introduction.tex
\section{Introduction}
\label{sec:intro}
Safety-critical software systems are those whose failure or malfunction will result in casualty and/or serious financial loss. Assessing and assuring safety is an important aspect of these kinds of systems. Over the past two decades, safety assurance cases (SACs) have emerged as a widely-used technique for safety justification and argument\cite{Bloomfield2010}. In contrast to the traditional prescriptive approach of safety control and certification, SACs emphasize proactive practices that rely on system developers taking the initiative and creating goal-oriented or claim-based safety arguments \cite{Hawkins2013}.

SAC arguments are often organized as a tree structure, usually graphical, that divides a top-level safety goal or claim into layers of arguments, which are eventually supported by safety evidence such as test reports or analysis results \cite{Bishop1998}. With such a structure, SACs are considered a useful technique to (1) help developers conduct safety management and (2) support regulatory experts to evaluate system safety \cite{UKMinistryofDefence2017, Graydon2017}. Several graphic notations were developed to help organize and represent SACs, including the Claims-Arguments-Evidence notation \cite{Adelard} and the Goal Structuring Notation \cite{Kelly2004}.

SACs has been recommended in many software-intensive safety-critical domains. In the US, the Food and Drug Administration (FDA) issued guidance that requests infusion pump manufactures to submit SACs as part of the safety approval process \cite{USFDA2014}. The Ministry of Defense of the UK requires all defense system contractors to provide SACs for their products and services \cite{UKMinistryofDefence2017}. Many international standards, including ISO26262 for road vehicles, IEC62425 for railway electronic systems, and IAEA SSG-23 for radioactive waste management systems, also recommended the use of SACs.

Despite, or rather because of the rapid growth, the SAC approach has suffered from several criticisms. Researchers and practitioners have criticized SACs because (1) its lack of guidance in constructing effective arguments \cite{Sujan2016}, (2) its tendency to suffer from confirmation bias \cite{Leveson2011}, (3) its reliance on the regulation culture to fulfill its value \cite{Steinzor2011}, and (4) its inefficacy to capture confidence and uncertainty issues \cite{Graydon2017, Duan2014}. More importantly, there is little knowledge regarding the perspectives of safety experts on the benefits, challenges, and best practices of using SACs. Research in safety-critical domains is usually system-centered, leaving the practitioners' values and perceptions even less investigated. The lack of this knowledge considerably diminishes effective adoption of the SAC approach in the development practice of safety-critical systems.

In this paper, we aimed to fill this gap by accumulating insights about the SAC approach from professional practitioners and safety experts who focused on safety-critical software systems. In particular, we conducted an empirical study that involved in-depth interviews with nine participants to identify (1) their perceived values and benefits of using SACs, (2) the challenges they met associated with SAC use, and (3) their insights into best practices and strategies for overcoming those challenges.


%% file: methods.tex
\section{Methods}
\label{sec:methods}
We recruited our participants through inviting personal contacts and industrial attendants of software safety conferences and workshops. We sent invitation emails to 39 practitioners and six academia researchers; the academia researchers were contacted because they have worked closely on industry-centered projects and had extensive experience creating industrial-grade SACs. Nine agreed to participate, including two software developers, two safety analysts, one research engineer, two system assessors from a US federal certifying agency, and two researchers from a US university.
Our participants covered various software-intensive safety-critical domains including automotive, railway systems, medical devices, and aviation systems. Participants' experience in the safety-critical systems field ranged from six to 25 years. Table \ref{tab:participantsSummary} summarizes our participants' professional experience.

\vspace{-4pt}
\begin{table}[ht]
\caption{Summary of participants' professional experience.}
\vspace{-6pt}
\centering
\renewcommand{\arraystretch}{0.8}
\small
\label{tab:participantsSummary}
\begin{tabular}{llc}

\toprule
\multicolumn{1}{c}{\textbf{ID}} & \multicolumn{1}{c}{\textbf{Role \& Job Title}} & \begin{tabular}[c]{@{}c@{}}\textbf{Years of  Experience}\end{tabular} \\

\midrule
P1 & Development: Senior software engineer & 8 \\
P2 & Development: Chief technology officer & 25 \\
P3 & Analysis: Safety specialist & 15 \\
P4 & Analysis: Functional safety specialist & 25 \\
P5 & Research: Safety technology researcher & 6 \\
P6 & Certifying: Senior systems engineer & 23 \\
P7 & Certifying: Software engineer & 10 \\
P8 & Academic Research: Project specialist & 17 \\
P9 & Academic Research: Scientist & 6 \\
\bottomrule
\end{tabular}
\vspace{-4pt}
\end{table}

We first conducted in-depth, semi-structured interviews with the nine participants. During the interviews, we asked the participants to describe (1) their experience in safety-critical systems and SACs, (2) their motivation and perceived benefits of using SACs, as well as (3) the challenges they faced and their mitigating strategies. Each interview took between 45 and 60 minutes. The interviews were audio-recorded and later fully transcribed.

Two researchers inductively coded the interviews to identify prominent themes in participants' consideration of the benefits, challenges, and best practices of using SACs. Upon reaching an agreement, a codebook was created \cite{miles2014qualitative}. One other researcher (a blind coder) then used it to deductively code the interviews \cite{miles2014qualitative}. We calculated inter-rater reliability using Cohen's Kappa through binary agreement between the codebook creators and the blind coder. We then refined the codebook by removing the codes that didn't achieve a ``substantial'' inter-rater agreement (i.e. kappa statistic lower than 0.6 \cite{Landis1977}).  Among all the themes that remained in our codebook, the average kappa statistic was .83 (SD = .16).

To further understand the participants' perceptions, we then asked them to complete a survey. In the survey, participants rated their perceived importance of the benefits and challenges we identified on a five-point Likert scale (from 1-Not at all important to 5-Extremely important) and provided additional comments about how to overcome those challenges. Following a grounded theory approach \cite{miles2014qualitative}, we inductively incorporated those comments in our codebook to refine our codes. Our interview and survey instruments are available at: \url{https://doi.org/10.6084/m9.figshare.5975446}.

%% file: results.tex
\section{Results}
\label{sec:results}
In this section, we describe our results from both the interviews and the surveys.

\subsection{Benefits}
When asked about the motivations and benefits of using SACs, our participants discussed factors that fell into one of the following four themes.
In the following sections, we report those themes and their average perceived importance (IMP) ratings from the survey.
\vspace{-4pt}
\subsubsection{Communication values}
($IMP=3.86$)
Seven participants mentioned that the clear structure of SACs supports communication of a safety argument among the development team and the safety assessors. For example, when asked about motivations of requiring manufactures to provide SACs for certification, a system assessor (P6) said, ``\textit{SACs are a good way for manufacturers to communicate with us and us with them. It's basically a means of organizing and presenting your design work to somebody who doesn't know anything about it and trying to convince them that it's safe and effective. ... Also, [when there is problem] we can be less ambiguous and more focused on what the problem is. That makes it easier for the manufacturers to understand what the regulators are looking for as well.}''

\vspace{-4pt}
\subsubsection{Frames thinking about system safety}
($IMP=3.86$)
Three participants, including both of the software developers, considered the process of creating and maintaining SACs as an effective framework for the development team to think about system safety in a more rigorous way. For example, a functional safety analyst (P4) mentioned, ``\textit{If you go with more effort and try to document the argument in SACs, you start realizing all the questions in your mind. You start realizing, `Oh. That's the assumption. But do we have evidence that it is true?' It forces you to be more rigorous in your thinking. ... All that the SAC did is trying to help you capture your thinking, your rationale, your understanding.}''

\vspace{-4pt}
\subsubsection{Fills the gap for new systems}
($IMP=3.71$)
Two of our participants, who were most experienced in SACs, also mentioned that the goal/claim-oriented SACs support safety argument for new systems, where safety regulations or guidelines do not yet exist. An academic researcher (P9) who created many SACs for industrial collaborators said, ``\textit{You can't regulate something that doesn't exist. So when the system is new and the product is novel, assurance cases fill the gap -- so that we can try to demonstrate that a new system is safe for use.}''

\vspace{-4pt}
\subsubsection{Easy to engage with}
($IMP=3.14$)
Four participants stated that the graphical representation of SACs is easy to understand and easy to engage with. For example, a system assessor (P7) said, ``\textit{The graphs are really easy to understand from a high level. ... It gives you an overview. You can easily build it in your brain.}'' A safety analyst (P4) also commented on straightforward notation of SACs: ``\textit{It's not too complicated. Especially with the Goal Structuring Notation, it isn't really got half a of dozen symbols. So it's quite easy to understand.}''

\subsection{Challenges}
\begin{figure}
\includegraphics[width=.46\textwidth]{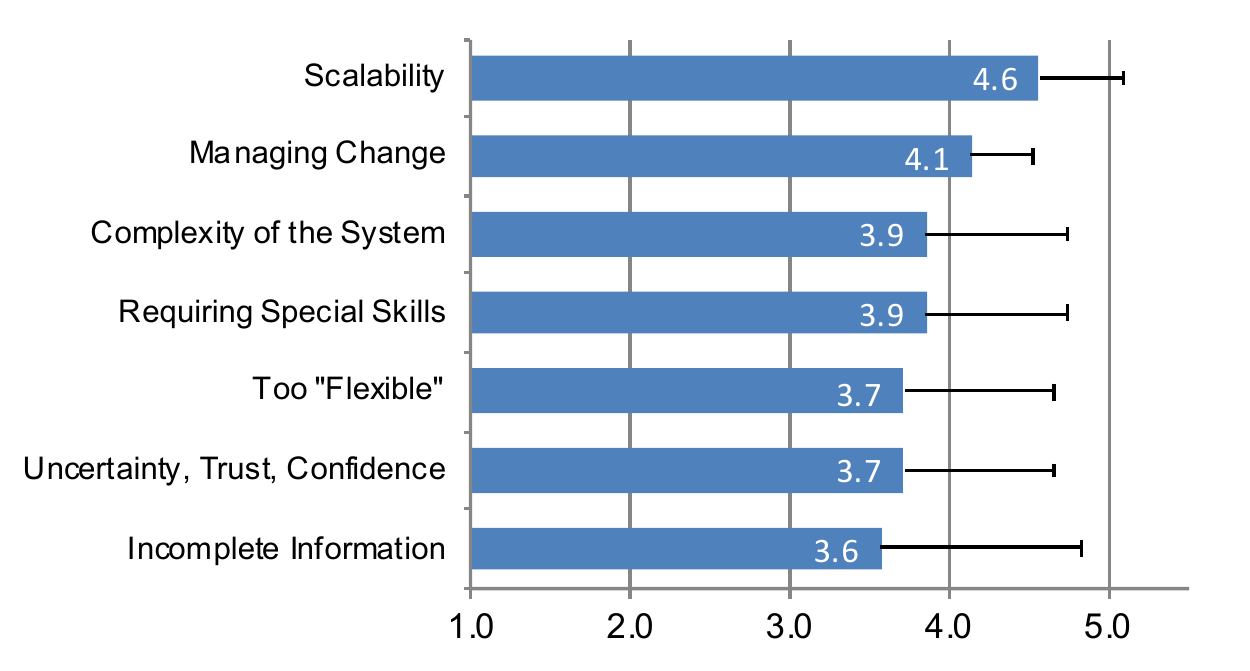}
\vspace{-6px}
\caption{Average perceived importance of the challenges in using SACs. Error bars show the standard deviations.}
\vspace{-6px}
\label{fig:ChallengeRatings}
\end{figure}

During the interviews, our participants described various factors that resulted in challenges of using SACs that we categorized into seven types. The survey results revealed that while varied, participants considered all of these types of challenges as at least moderately important; all average ratings are above three (see Fig. \ref{fig:ChallengeRatings}). In the following sections, we describe these challenges in the descending order of average importance as rated by the participants.

\subsubsection{Scalability}
The most significant challenge from our participants' perspective is that of navigating and comprehending large SACs, especially when presented using graphical form. Our participants mentioned that this issue is exacerbated by the limited tool support for creating, maintaining, and reviewing SACs. While there are a few dedicated commercial SAC tools such as Adelard's ASCE (www.adelard.com/asce) and GessNet's TurboAC (www.gessnet.com), current tools either fail to deliver features for effectively handling scalability or are focused on a narrow domain with limited adoption.

\vspace{-4pt}
\subsubsection{Managing change}
Changes in software, especially in software requirements, can result in changes in SACs. But there is currently no effective mechanisms to manage those changes. Our participants mentioned that this is particularly challenging because the SAC creation and maintenance has not been fully integrated into the software development process. The lack of tool support also contributed to this challenge.

\vspace{-4pt}
\subsubsection{Requiring special skills to create}
The graphic notations of SACs are usually easy to understand. However, our participants considered that creating a convincing, well structured safety argument requires special skills and considerable experience. Participants also mentioned that because of sensitivity of many safety-critical domains, there is limited knowledge sharing about practical SAC strategies and real-world SAC examples, making acquiring those skills even more difficult.

\vspace{-4pt}
\subsubsection{Complexity of the system}
Our participants mentioned that since most safety-critical systems are innately complex and many systems are interconnected, capturing the safety concerns in those systems becomes increasingly challenging. In addition, because developing safety-critical systems often involves experts in various disciplines, multidisciplinary collaboration in arguing system safety is also important and challenging.

\vspace{-4pt}
\subsubsection{Uncertainty, trust, confidence}
When considering the efficacy of SACs, many participants voiced concerns about the fact that system safety always involves issues related to uncertainty, trust, and confidence. Capturing these ``intangible'' issues and establishing trust and confidence in the safety arguments was considered as a challenge.

\vspace{-4pt}
\subsubsection{Too "flexible"}
SACs rely on system manufacturers to create safety justifications and arguments. While this mechanism offers flexibility, some of our participants mentioned that it is dangerous if manufacturers overlook or omit certain safety aspects. In other words, the SAC technique may be subject to confirmation bias and/or conflicts of interest of the manufacturers. Related to the challenge of establishing trust and confidence, participants considered addressing these limitations of SACs challenging.

\vspace{-4pt}
\subsubsection{Incomplete information}
Related to the limited integration of SAC management in the software development process, our participants discussed the challenges involved in gathering sufficient and accurate information for safety arguments. This issue primarily originated from flawed safety requirements, insufficient test coverage, and incomplete traceability across software artifacts.

\subsection{Best Practices}
During the interviews and in the survey responses, our participants provided useful insights into strategies addressing the challenges of using SACs. While a few targeted a specific challenge, most of those strategies were considered as general best practices for creating and maintaining good safety arguments using SACs.

\vspace{-4pt}
\subsubsection{Start early and update often}
Seven of our participants mentioned that it is best to start creating the initial SAC early in the software development process and let the SAC evolve with other software artifacts. This usually leverages and also reinforces software traceability. For example, a safety analyst (P3) discussed his successful experience of creating a SAC for an aerospace system, saying, ``\textit{We started [building the SAC] from the concept phase down to the specification phase, the architecture and so on. And we tried to trace everything that comes from the concept and specifications down to at least the test case, and the test report.}'' An academic researcher (P8) considered that updating SACs along with system development is the key to fully realizing the benefits of SACs: ``\textit{Essentially constructing the safety case should drive design choices and decisions that support system safety}''.

\vspace{-4pt}
\subsubsection{Strengthen analysis process}
Seven participants also suggested that practitioners should strengthen their safety analysis process by involving multiple stakeholders, including external reviewers, and strive for comprehensiveness of safety requirements. For example, a developer (P1) considered that to ``\textit{have a diverse team, especially with people who are not software developers, discuss the SAC and come to an agreement}'' is important to raise the team's confidence on the safety justification and argument.

\vspace{-4pt}
\subsubsection{Focus on direct and well-structured argument}
Six participants discussed the importance of focusing on creating direct, defensible, and clearly divided arguments, supported by relevant and up-to-date evidence. For example, when asked about suggestions of creating a good SAC, a safety assessor (P7) said, ``\textit{If you think about claim, argument and evidence, I want to see those three things come together. The structure of the claims should be very logically divided. ... And you need a defensible argument and tailored evidence.}'' A software developer (P2) also discussed his strategy to ``orthogonaly'' divide and conquer safety concerns in order to achieve a well-structured argument.

\vspace{-4pt}
\subsubsection{Augment graphs with text and tables}
The graphic representation could support the comprehensibility of SACs. However, it is still important to combine the graphs with text and tables to augment the argument. Five of our participants discussed this issue. For example, a safety analyst (P4) said, ``\textit{I tend to use the Goal Structuring Notation in small amounts. And then put it in the context of text and tables. So you tend to have text to talk about stuff and then say, 'Okay. This is an overall argument. And here is a diagram.' ... So later when I read the texts and the tables, I got a context from the standing in the diagram -- sort of why they are there and what they are supposed to be justifying.}''

\vspace{-4pt}
\subsubsection{Seek for tool support}
While many available SAC tools have limitations, six participants mentioned that it is still worth seeking tool support to address scalability issues and change impact analysis in SACs. Three participants have used internally developed tools to create and manage SACs. The two safety assessors also mentioned that they found that manufactures started to submit SACs developed using commercial tools; SACs submitted with tools usually eased their reviewing process. We believe the availability and maturity of SAC tools will go hand in hand with the adoption of the SAC technique.

\vspace{-4pt}
\subsubsection{Promote knowledge sharing}
Four of our participants discussed the importance for practitioners and researchers in the safety community to promote knowledge sharing about SACs by publishing successful real-world examples and developing standards, patterns and guidelines. Particularly, a safety analyst (P4) mentioned he benefited a lot from using SAC patterns \cite{Hawkins2009, Denney2016}. An academic researcher (P8) also expressed his hope to see wider adoption and more knowledge sharing: ``\textit{Part of the issue is that the area itself is not well-developed and the techniques are not widely-practiced. As more system engineers and safety engineers develop safety-cases and share best practices, the problems would get addressed.}''

%% file: conclusion.tex
\section{Discussion and Conclusion}
\label{sec:conclusion}
In this study, we investigated the perspectives of practitioners and safety experts on using SACs in safety-critical domains. In general, our participants found the SAC approach beneficial for communicating safety arguments and managing safety issues. Particularly, a general theme of those benefits lies on the SACs' ability to bridge the expertise and work-flows of software developers, safety analysts, and product certification experts. In other words, our participants considered SACs to be a useful tool for supporting collaboration of experts from different areas to enhance system safety.

However, because the SAC is still an emerging technique, our participants faced various challenges when using SACs. Our study revealed that those challenges were mostly associated with (1) the lack of tool support (to address scalability issues and support change management), (2) insufficient process integration (so that change management tended to be ad hoc and gathering accurate safety information was difficult), and (3) scarcity of experienced personnel (to reliably capture safety-related issues in complex systems). 

Our participants collectively offered best practices and strategies to overcome those challenges, which can be seen on three different levels. First, on a tactical level, focusing on creating a direct argument structure and augmenting the graphical representation would help create cogent and effective SACs. Second, on a process and organizational level, integrating the SAC approach early in the development process and creating a culture of 
involving diverse experts in safety analysis would strengthen safety justification. In fact, this is a strategy repeatedly discussed in the safety literature (e.g. \cite{Bishop1998, UKMinistryofDefence2017, Sujan2016}) but rarely practiced in the real world. Finally, effective adoption of SACs requires community efforts to promote tool support and knowledge sharing. With more organizations realizing the value of this technique and more regulatory bodies requiring SACs, we expect to see an accelerated growth of work in this area.

This paper is one of the first to capture the perspectives of practitioners and safety experts on the use of SACs. We were only able to work with nine participants. However, they covered a wide range of safety-critical domains and included experts with different roles associated with the use of the SAC approach. From this diverse body of participants, we aimed to identify common elements in their perceptions and opinions. We argue that these common elements captured representative issues of SAC use and served as a much-needed first step towards knowledge sharing in this area.